\documentclass[11pt]{revtex4}
\usepackage{amssymb}
\usepackage{latexsym}   %Revised in June 6, 2016 at ICTP%
\usepackage{epsfig}
\begin{document}

\title{Effects of backreaction and exponential nonlinear electrodynamics on the holographic superconductors}
\author{A. Sheykhi$^{1,2}$\footnote{asheykhi@shirazu.ac.ir},  F. Shaker $^{1}$}
\address{$^1$ Physics Department and Biruni Observatory, College of
Sciences, Shiraz University, Shiraz 71454, Iran\\
$^2$ Research Institute for Astronomy and Astrophysics of Maragha
(RIAAM), P.O. Box 55134-441, Maragha, Iran}

\begin{abstract}
We analytically study the properties of a $(2+1)$-dimensional
$s$-wave holographic superconductor in the presence of exponential
nonlinear electrodynamics. We consider the case in which the
scalar and gauge fields back react on the background metric.
Employing the analytical Sturm-Liouville method, we find that in
the black hole background, the nonlinear electrodynamics
correction will affect the properties of the holographic
superconductors. We find that with increasing both backreaction
and nonlinear parameters, the scalar hair condensation on the
boundary will develop more difficult. We obtain the relation
connecting the critical temperature with the charge density. Our
analytical results support that, even in the presence of the
nonlinear electrodynamics and backreaction, the phase transition
for the holographic superconductor still belongs to the second
order and the critical exponent of the system always takes the
mean-field value $1/2$.

\end{abstract}
\maketitle

\section{Introduction}
The AdS/CFT correspondence is an equivalence between a conformal
field theory (CFT) in $d$ spacetime dimensions, and a theory of
gravity in $(d+1)$-dimensional anti-de Sitter (AdS) spacetime
\cite{AdS.CFT1, AdS.CFT2, AdS.CFT3}. The $d$-dimensional theory
does not have a gravitational force, and is to be viewed as a
\textit{hologram} of the $(d+1)$-dimensional theory. The AdS/CFT
correspondence is a well-known approach to explore strongly
coupled field theories in which certain questions become
computationally smooth and conceptually more explicit. The AdS/CFT
correspondence can be applied to condensed matter phenomena. In
condensed matter physics, there are many strongly coupled systems
such as superconductors. In this regards, it was recently argued
that it is quite possible to shed some light on the problem of
understanding the mechanism of the high temperature
superconductors in condensed matter physics, by studying a
classical general relativity in one higher dimensional spacetime
\cite{Har1,Har2}. The holographic superconductivity is a
phenomenon associated with asymptotic AdS black holes. The studies
on the holographic superconductors have received a lot of
attentions \cite{ HS.HM, HS.H,HS.FGR,Wang1,Wang2}.

Most studies on the holographic superconductors are focused on the
cases where the gauge field is in the form of the linear Maxwell
field. But nonlinear electrodynamics is constructed by the desire
to find non-singular field theories. One may consider nonlinear
electrodynamics as a possible mechanism for avoiding the
singularity of the point-like charged particle at the origin. The
nonlinear extension of the original Maxwell electrodynamics in the
context of holographic superconductors have arisen intensive
investigations \cite{b.Nu.GB.BI, p.BI&LN.4D, p.BI.4D,
JC.p.Nu.BI.4D, P.BGRL,P.JPC, b.BI.5D,Shey2}. In particular, in
order to see what difference will appear for holographic
superconductor in the presence of Born-Infeld (BI) nonlinear
electrodynamics, compared with the case of linear Maxwell
electrodynamics, the authors of Ref. \cite{LPJW2015.p.BI} have
studied condensation and critical phenomena of the holographic
superconductors with BI electrodynamics in $d$-dimensional
spacetime. Their analytical results indicate that the nonlinear BI
electrodynamics decreases the critical temperature of the
holographic superconductor. It was observed that the higher BI
corrections make it harder for the condensation to form but do not
affect the critical phenomena of the system \cite{LPJW2015.p.BI}.

It is also interesting to investigate the effects of gauge and
scalar fields of the holographic superconductor on the background
geometry. Although if we ignore this backreaction, the problem is
simplified, but retains most of the interesting physics since the
nonlinear interactions are retained. Indeed, considering the
holographic superconductor model away from the probe limit may
bring rich physics. Therefore, many authors have tried to study
the holographic superconductors away from the probe limit
\cite{PJWC2012, B1, B2, B3, B4, B5, B6}. Employing the analytical
Sturm-Liouville  method, the effects of both backreaction and BI
nonlinear parameter on the critical temperature as well as scalar
condensation were explored in Ref. \cite{BIBR}. Furthermore,  the
relation between the critical temperature and charge density was
established \cite{BIBR}. It was shown that it is more difficult to
have scalar condensation in BI electrodynamics when the
backreaction is taken into account \cite{BIBR}.

In the present work we would like to extend our analytical study
on the backreacting holographic superconductors by considering
another form of the higher order corrections to the gauge field,
i.e., the exponential form of nonlinear electrodynamics. It was
shown that when the backreaction is taken into account, even the
uncharged scalar field can form a condensation in the
$(2+1)$-dimensional holographic superconductor model \cite{Har2}.
Numerical studies on the holographic superconductors with
exponential nonlinear (EN) electrodynamics are carried out in the
probe limit \cite{ZPCJ.P.N.N}. It was shown that the higher
nonlinear electrodynamics corrections makes the condensation
harder to form \cite{ZPCJ.P.N.N}. As far as we know, analytical
study on the holographic superconductor in the presence of EN
electrodynamics and away from the probe limit has not been done.
Considering exponential form of the higher corrections to the
gauge field, we shall analytically investigate the properties of
the holographic superconductors when the gauge and scalar field do
back react on the metric background. We shall use the analytical
Sturm-Liouville eigenvalue problem. We will also compare our
results with those for the holographic superconductors with BI
nonlinear electrodynamics with backreaction given in \cite{BIBR}.

This paper is outlined as follows. In section \ref{Basic}, we
introduce the action and basic field equations of the
$(2+1)$-holographic superconductor with EN electrodynamics with
backreaction. In section \ref{Field}, we compute the critical
temperature in terms of the charge density and disclose its
dependence on the both nonlinear and backreaction parameters.
Section \ref{Cri}, includes step-by-step computations for
obtaining the critical exponent and the condensation values of the
holographic superconductor and provides explanations about them.
Section \ref{Con} will help us to collect the obtained results
briefly.

%%%%%%%%%%%%%%%%%%%%%%%%%%%%%%%%%%%%%%%%%%%%%%%%%%%%%%%%%%%%%%
\section{Basic Equations of Holographic Superconductors with Backreactions \label{Basic}}
The action of Einstein gravity coupled to a charged complex scalar
field in the presence of nonlinear electrodynamics is described by
\begin{eqnarray}\label{action}
S=\int d^{4}x \sqrt{-g}\left[\frac{1}{2 \kappa^{2}}( R-2 \Lambda)
+\mathcal{L}(\mathcal{F})- | \nabla\psi - i q A \psi|^{2} - m^{2}
|\psi|^{2}\right],
\end{eqnarray}
where $\kappa$ is the usual four dimensional gravitational
constant, $\kappa^2=8\pi G_{4}$, $\Lambda=-{3}/{L^2}$ is the
cosmological constant, where $L$ is the AdS radius which will be
scaled unity in our calculations. $R$ and $g$ are, respectively,
representing the Ricci scalar and the determinant of the metric.
$A$ is the gauge field and $\psi$ represents a scalar field with
charge $q$ and mass $m$. $\mathcal{L}(\mathcal{F})$ is a simple
generalization of Maxwell Lagrangian in a exponential form
\cite{Hendi}
 \begin{equation}
\mathcal{L}(\mathcal{F})= \frac{1}{4b} \left(e^{-b \mathcal
{F}}-1\right),
\end{equation}
where $b$ is the nonlinear parameter, $\mathcal
{F}=F_{\mu\nu}F^{\mu\nu}$ and $F^{\mu\nu}$ is the electromagnetic
field tensor. Expanding this nonlinear Lagrangian for small $b$,
the leading order term is the linear Maxwell theory,
$\mathcal{L}(\mathcal {F})=-\mathcal {F}/{4}+\mathcal{O}(b)$. The
plane-symmetric black hole solution with an asymptotically AdS
behavior including the backreaction is described by the metric,
\begin{eqnarray}\label{metric}
ds^{2}&=- f(r) e^{-\chi(r)} dt^{2}+\frac{dr^2}{f(r)}+r^{2}(dx^2 +
dy^2).
\end{eqnarray}
We adopt the following gauge choices for the vector field and the
scalar field,
\begin{eqnarray}\label{phipsi}
A_{\mu}=\left(\phi(r),0,0,0\right),    \  \   \   \psi=\psi(r),
\end{eqnarray}
with these functions being real-valued. Then, we need to establish
the Einstein equations  by varying action (\ref{action}) with
respect to the metric. We find
\begin{eqnarray}\label{Eq1}
R^{\mu\nu}-\frac{g^{\mu\nu}}{2} R -\frac{3 }{L^2}g^{\mu\nu}=
\kappa^2 T^{\mu\nu},
\end{eqnarray}
where the energy momentum tensor is given by
\begin{eqnarray}\label{Eq1T}
T^{\mu\nu}&=&\frac{1}{4b} g^{\mu\nu}
\left(e^{-b\mathcal{F}}-1\right)+ e^{-b\mathcal{F}}F_{\sigma}^{ \
\mu} F^{\sigma\nu}-m^2 g^{\mu\nu}
|\psi|^2-g^{\mu\nu} | \nabla\psi - i q A \psi|^{2}\nonumber\\
&& +\left[(\nabla^{\nu}+ i q A^{\nu}) \psi^{*}  (\nabla^{\mu}- i q
A^{\mu})\psi+\mu \leftrightarrow\nu\right].
\end{eqnarray}
Variation with respect to the scalar field yields
\begin{eqnarray}\label{Eq2}
(\nabla_{\mu} -i q A_{\mu}) (\nabla^{\mu}- i q A^{\mu}) \psi -m^2
\psi =0,
\end{eqnarray}
while the electrodynamic equation,
\begin{eqnarray}\label{Eq3}
\nabla_{\mu} \left(F^{\mu\nu}e^{-b\mathcal{F}} \right)= i q \Bigg
[ \psi^{*} (\nabla^{\mu}- i q A^{\mu})\psi -\psi (\nabla^{\mu}+ i
q A^{\mu}) \psi^{*}\Bigg],
\end{eqnarray}
is obtained by varying action (\ref{action}) with respect to the
gauge field. These equations can easily reduce to those of the
holographic superconductor in Maxwell theory \cite{Har2}, provided
$b\rightarrow0$. Calculations of the Einstein, scalar and
electrodynamic field equations, with respect to the metric
(\ref{metric}), yield the following expressions,
\begin{eqnarray}
\chi'&+2 r \kappa^{2} \left(\psi'^{2}+\frac{q^{2}e^{\chi} \phi^{2}
\psi^2}{f^2}\right)=0, \label{Eqchi}
\end{eqnarray}
\begin{eqnarray}
f^{\prime}-\left(\frac{3 r}{L^{2}}-\frac{f}{r}\right)-\frac{\chi'}{2} f+ r\kappa^{2}
\left[m^{2} \psi^{2}+\frac{1}{4b}(1-e^{2b\phi '^{2}e^{\chi}})+\phi '^{2}e^{\chi+2b\phi '^{2}e^{\chi}}\right]=0,\label{Eqf}
\end{eqnarray}
\begin{eqnarray}\label{Phir}
\phi''(1+4 b e^{\chi}\phi'^2)+\frac{2}{r}\phi' (1+r be^{\chi}\chi' \phi '^{2})+\frac{\chi' \phi '}{2}-\frac{2 q^2
\phi\psi^2}{f} e^{-2be^{\chi}\phi'^2} =0,
\end{eqnarray}
\begin{eqnarray}\label{psir}
\psi''+\left(\frac{f'}{f}+\frac{2}{r}-\frac{\chi'}{2}\right)\psi'+\left(\frac{q^2
e^{\chi}\phi^2}{f^2}-\frac{m^2}{f}\right)\psi=0,
\end{eqnarray}
where the prime denotes derivative with respect to $r$. We further
assume there exists an event horizon $r_{+}$  for which
$f(r_{+})=0$, and thus the corresponding Hawking temperature of
the black hole reads
\begin{eqnarray}\label{T}
T=\frac{f'(r_{+}) e^{-\chi(r_{+})/2}}{4\pi}.
\end{eqnarray}
For the case with $b\rightarrow0$, Eqs. (\ref{Eqchi})-(\ref{psir})
coincide with their corresponding equations presented in
\cite{PJWC2012}. Also in the probe limit where $\kappa=0$, Eqs.
(\ref{Phir}) and (\ref{psir}) go back to the $(2+1)$-dimensional
holographic superconductor model studied in \cite{ZPCJ.P.N.N}. In
this case the solution of Eq. (\ref{Eqf}) is
\begin{eqnarray}\label{f1}
f(r)=\frac{r^2}{L^2}\left(1-\frac{r_{+}^3}{r^3}\right).
\end{eqnarray}
It should be noted that we can set the charge parameter, $q$, as
unity and keep $\kappa^2$ finite when the backreaction is taken
into account by adopting the scaling symmetry \cite{ssym}. When
the Hawking temperature is above the critical temperature
$T>T_{c}$, the system leads to the well-known exact black holes as
$b\rightarrow0$ with the metric coefficient and the potential
function given by
\begin{eqnarray} \label{frphr}
f(r)=\frac{r^2}{L^2}-\frac{1}{r}
\left(\frac{r_{+}^3}{L^2}+\frac{\kappa^2 \rho^2}{2
r_{+}}\right)+\frac{\kappa^2 \rho^2}{2 r^2},  \   \   \
\phi\approx\mu-\frac{\rho}{r}.
\end{eqnarray}
On the dual side, $\mu$ and $\rho$ are, respectively, the chemical
potential and charge density of the holographic superconductor.
When $\kappa=0$, the metric coefficient $f(r)$ recovers the case
of Schwarzschild AdS black holes (\ref{f1}). For investigating the
properties of dual model in superconducting phase, i.e.,
$\psi(r)\neq0$, we need the suitable boundary conditions.
Examining the behavior of the fields near the horizon, we find the
suitable boundary conditions as
\begin{eqnarray}
\phi(r_{+})=0,\  \   \
\psi(r_{+})=\frac{f'(r_{+})\psi'(r_{+})}{m^2},
\end{eqnarray}
and hence the metric functions $\chi$ and $f(r)$ satisfy
\begin{eqnarray}
\chi'(r_{+})&=-2 r_{+} \kappa^{2} \left(\psi'^{2}(r_{+})+\frac{q^{2}e^{\chi(r_{+})} \phi'^{2}(r_{+})
\psi^2(r_{+})}{f'^2(r_{+})}\right),
\end{eqnarray}
\begin{eqnarray}\label{f+}
f^{\prime}(r_{+})=\frac{3 r_{+}}{L^{2}}- r_{+}\kappa^{2}
\left[m^{2} \psi^{2}(r_{+})+\frac{1}{4b}(1-e^{2b\phi
'^{2}(r_{+})e^{\chi(r_{+})}})+\phi
'^{2}(r_{+})e^{\chi(r_{+})+2b\phi
'^{2}(r_{+})e^{\chi(r_{+})}}\right].
\end{eqnarray}
The asymptotic behavior of the fields, corresponding to the
solution of Eqs. (\ref{Phir}) and (\ref{psir}) in the limit
$r\rightarrow\infty$, are given by
\begin{eqnarray}\label{BC.Phi}
&&\phi\approx\mu-\frac{\rho}{r},\\
&&\psi\approx\frac{\psi_{-}}{r^{\Delta_{-}}}+\frac{\psi_{+}}{r^{\Delta_{+}}},
\end{eqnarray}
where
\begin{eqnarray}
\Delta_{\pm}=\frac{3}{2}\pm\frac{\sqrt{9+4 m^2}}{2},
\end{eqnarray}
is the conformal dimension of the dual operator
$\mathcal{O_{\pm}}$ in the boundary field theory. Here $\psi_{+}$
and $\psi_{-}$ can be considered as the source and the vacuum
expectation values of the dual operator. Hereafter, we set
$\psi_{+}=0$ and investigate the condensation of
$\psi_{-}=<\mathcal{O_{-}}>$, analytically. In what follows we
choose the scalar to have $m^2=-2$, and hence the corresponding
dual operator has mass dimension $\Delta_{-}=1$.
%%%%%%%%%%%%%%%%%%%%%%%%%%%%%%%%%%%%%%%%%%%%%%%%%%%%%%%%%%%%%%%%%%
\section{Analytical study and critical tempreture\label{Field}}
In this section, we investigate the analytical properties of a
$(2+1)$-holographic superconductor in the framework of EN
electrodynamics. We study the problem by taking the backreaction
into account. We find the critical temperature $T_{c}$ via the
Sturm-Liouville variational approach. Further, we obtain a
relationship between the critical temperature and the charge
density and investigate the effects of both backreaction and EN
parameter on the critical temperature. In order to get the
solutions in superconducting phase, we can define a new variable
$z={r_{+}}/{r}$. Then, the equations of motion can be rewritten as
\begin{eqnarray}\label{chiz}
\chi^{\prime}-2 \kappa^{2}\left(z
{\psi^{\prime}}^{2}+\frac{r_{+}^{2}}{z^{3} f^{2}} e^{\chi}
\phi^{2} \psi^{2}\right)=0,
\end{eqnarray}
\begin{eqnarray}\label{fz}
f^{\prime}-\frac{f}{z} +\frac{3 r_{+}^{2}}{z^{3}}
-\frac{\chi'f}{2}- \frac{\kappa^{2} r_{+}^{2}}{z^{3}}\left[m^{2}
\psi^{2} +\frac{1}{4b} (1-e^{\frac{2bz^4}{r_{+}^2}\phi
'^{2}e^{\chi}})+\frac{z^4}{r_{+}^2}\phi'^{2}e^{\chi+\frac{2bz^4}{r_{+}^2}\phi
'^{2}e^{\chi}} \right]=0,
\end{eqnarray}
\begin{eqnarray}\label{Phiz}
\phi''\Bigg(1+ \frac{4bz^4}{r_{+}^2}e^{\chi}\phi'^2
\Bigg)+\frac{8bz^3}{r^2_{+}}e^{\chi}\phi'^3
+\frac{2bz^4}{r_{+}^2}e^{\chi}\phi'^3 \chi'+\frac{\phi'\chi'}{2}
-\frac{2r_{+}^2\psi^2}{fz^4}e^{-\frac{2bz^4}{r^2_{+}}e^{\chi}\phi'^2}
\phi=0,
\end{eqnarray}
\begin{eqnarray}\label{psiz}
\psi''-\left(\frac{\chi'}{2}-\frac{f'}{f}\right)\psi'-\frac{r_{+}^2}{z^4}
\left(\frac{m^2}{f}-\frac{e^{\chi}\phi^2}{f^2}\right)\psi=0,
\end{eqnarray}
where now the prime denotes derivative with respect to $z$. When
$b\rightarrow0$, the above equations restore the corresponding
equations in Ref. \cite{PJWC2012}, while in the absence of the
backreaction, Eqs. (\ref{Phiz}) and (\ref{psiz}) reduce to their
corresponding equations in Ref. \cite{ZPCJ.P.N.N}. Following the
perturbation scheme, since close to the critical point, the value
of the scalar operator is small, it can be introduced as an
expansion parameter
\begin{eqnarray}
\epsilon\equiv <\mathcal{O}_{i}>,
\end{eqnarray}
with $i=+$ or $i=-$. Besides, near the critical point the scalar
and gauge fields are small and therefore we can expand the gauge
field $\phi$, the scalar field $\psi$, and the metric functions
$f(z)$, $\chi(z)$ as \cite{PJWC2012}
\begin{eqnarray}
\psi=\epsilon \psi_{1}+\epsilon^3 \psi_{3}+\epsilon^5\psi_{5}+...,
\end{eqnarray}
\begin{eqnarray}
\phi=\phi_{0}+\epsilon^2\phi_{2}+\epsilon^4\phi_{4}+...,
\end{eqnarray}
\begin{eqnarray}
f=f_{0}+\epsilon^2 f_{2}+\epsilon^4f_{4}+...,
\end{eqnarray}
\begin{eqnarray}
\chi=\epsilon^2\chi_{2}+\epsilon^4\chi_{4}+...,
\end{eqnarray}
where the metric function $f(z)$ and $\chi(z)$ are expanded around
the Reissner-Nordstr\"{o}m AdS spacetime. Also, the chemical
potential $\mu$ may be expanded as \cite{PJWC2012}
\begin{eqnarray}
\mu=\mu_{0}+\epsilon^2\delta \mu_{2}+...,
\end{eqnarray}
where $\delta \mu_{2}$ is positive. Thus, near the phase
transition, the order parameter as a function of the chemical
potential has the form
\begin{eqnarray}
\epsilon\thickapprox\Bigg(\frac{\mu-\mu_{0}}{\delta
\mu_{2}}\Bigg)^{1/2}.
\end{eqnarray}
whose critical exponent $\beta=1/2$ is the same as in the
Ginzburg-Landau mean field theory. The phase transition can take
place when $\mu\rightarrow\mu_{0}$. In this case the critical
value of the chemical potential is given by $\mu_{c}=\mu_{0}$.

From Eq. (\ref{Phiz}) the equation for $\phi$ is obtained at
zeroth order as
\begin{eqnarray}\label{Phi0}
\phi''(z)\Bigg(1+4b \frac{z^4}{r_{+c}^2}\phi'^2
\Bigg)+\frac{8bz^3}{r^2_{+c}}\phi'^3(z)=0,
\end{eqnarray}
which admits the following solutions for the gauge field
\begin{eqnarray}\label{Phibeta1}
\phi(z)=\int_{1}^{z}{dz \frac{r_{+c}}{2z^2
\sqrt{b}}\sqrt{L_{W}\left(\frac{4bz^4 \beta^2}{r_{+c}^2}\right)}}.
\end{eqnarray}
In the above expression $\beta$ is an integration constant and
$L_W(x)={LambertW(x)}$ is the Lambert function which satisfies
\cite{Lambert}
\begin{equation}
L_W(x)e^{L_W(x)}=x,
\end{equation}
and has the following series expansion
\begin{equation}\label{LW}
L_W(x)=x-x^2+\frac{3}{2}x^3-\frac{8}{3}x^4+....
\end{equation}
Obviously, the series (\ref{LW}) converges provided $|x| <1$. If
we expand the solution (\ref{Phibeta1}) for small $b$ and keep the
only linear terms in $b$, we arrive at
\begin{eqnarray}\label{Phibeta}
\phi(z)=-\beta(1-z)+\frac{2\beta^3 b}{5r_{+c}^2}
\left(1-z^5\right)+\mathcal{ O} (b^2).
\end{eqnarray}
Differentiating Eqs. (\ref{BC.Phi}) and (\ref{Phibeta}) with
respect to $z$ and equating them at $z=0$, we find
$\beta=-{\rho}/{r_{+c}}$. Rearranging Eq. (\ref{Phibeta}) and
using the relation $\beta$, we arrive at
\begin{eqnarray}\label{Phi}
\phi_{0}(z)=\lambda
r_{+c}(1-z)\Bigg\{1-\frac{2}{5}b\lambda^2(1+z+z^2+z^3+z^4)\Bigg\},
\  \   \   b\lambda^2<1,
\end{eqnarray}
where
\begin{eqnarray}\label{lambda}
\lambda=\frac{\rho}{r^2_{+c}},
\end{eqnarray}
and we have neglected $\mathcal{ O} (b^2)$. Thus, to zeroth order
the equation for $f$ is solved as
\begin{eqnarray}
f_{0}(z) = r_{+} ^2 g(z)= r_{+} ^2 \left[ \frac{1}{z^2} -z -
\frac{\kappa ^2 \lambda^2}{2} z(1-z) + \frac{b}{10} \kappa^2
\lambda^4 z\left(1-z^5\right)\right].
\end{eqnarray}
At the first order, the behavior of $\psi$ at the asymptotic AdS
boundary is given by
\begin{eqnarray}
\psi_{1} \approx \frac{\psi_{-}}{r_{+}^{\Delta_{-}}}
z^{\Delta_{-}}+ \frac{\psi_{+}}{r_{+}^{\Delta_{+}}}z^{\Delta_{+}}.
\end{eqnarray}
Next, we introduce a variational trial function $F(z)$ near the
boundary
\begin{eqnarray}\label{psi1F}
\psi_{1}(z)= \frac{<\mathcal{O}_i>}{ \sqrt{2
}r_{+}^{\triangle_{i}}} z^ {\triangle_{i}}F(z),
\end{eqnarray}
with the boundary condition $F(0)=1$ and $F'(0)=0$. Then, we can
obtain the equation of motion for $F(z)$ by substituting
(\ref{psi1F}) into Eq. (\ref{psiz}). We find
\begin{eqnarray}\label{FF}
&&F^{\prime\prime}(z)+\Bigg[\frac{2\Delta}{z}+\frac{g'}{g}\Bigg]
F'(z)+\Bigg[\frac{\Delta}{z}\Bigg(\frac{\Delta-1}{z}+
\frac{g'}{g}\Bigg)-\frac{m^2}{z^4g}\Bigg]
F(z)\nonumber\\&&+\frac{\lambda^2 (1-z)^2}{z^4 g^2}\Bigg[1-\frac{4}{5}b
\lambda^2 \Bigg(1+z+z^2+z^3+z^4\Bigg)\Bigg]F(z)=0.
\end{eqnarray}
Defining the new functions
\begin{eqnarray}
T(z)&=&z^{2 \Delta_{i}+1} \Bigg[2(z^{-3} -1)-\kappa^2
\lambda^2(1-z)+\frac{b}{5}\kappa^2 \lambda^4 (1-z^5)\Bigg],\\
P(z)&=& \frac{\Delta_{i}}{z}\Bigg(\frac{\Delta_{i}-1}{z}+ \frac{g'}{g}\Bigg)-\frac{m^2}{z^4g},\\
Q(z)&=& \frac{ (1-z)^2}{z^4 g^2}\Bigg[1- \frac{4}{5}b \lambda^2
\Bigg(1+z+z^2+z^3+z^4\Bigg)\Bigg].
\end{eqnarray}
we can rewrite Eq. (\ref{FF}) as
\begin{eqnarray}
T F^{\prime\prime}+T' F' + P F +\lambda^2 Q F=0.
\end{eqnarray}
According to the Sturm-Liouville eigenvalue problem \cite{Gan},
the eigenvalue $\lambda^2$ can be obtained by minimizing the
expression
\begin{eqnarray}\label{lambda2}
\lambda^2=\frac{\int_{0}^{1} T\left(F'^2-P
F^2\right)dz}{\int_{0}^{1} TQF^2 dz},
\end{eqnarray}
where we have chosen the trial function in the form $F(z)=1-\alpha
z^2$. In order to simplify our calculations, we express the
backreaction parameter as \cite{PJWC2012}
\begin{eqnarray}
\kappa_{n}= n \Delta\kappa,   \    \   \    n=0,1,2,...
\end{eqnarray}
where $\Delta\kappa=\kappa_{n+1} - \kappa_{n}$ is the step size of
our iterative procedure. The main purpose is to work in the small
backreaction approximation so that all the functions can be
expanded by $\kappa^{2}$ and the $\kappa^{4}$ term can be
neglected. Furthermore, we retain the terms that are linear in
nonlinear parameter $b$ and keep terms upto $\mathcal{O}(b)$. So
we use the following relations
\begin{eqnarray}
\kappa^2 \lambda^2= \kappa_{n} ^2 \lambda^2=\kappa_{n} ^2
(\lambda^2|_{\kappa_{n-1}})
+\mathcal{O}\left[(\Delta\kappa)^4\right],
\end{eqnarray}
\begin{eqnarray}
b \lambda^2=b \left(\lambda^2|_{b=0}\right) +\mathcal{O} (b^2),
\end{eqnarray}
and
\begin{eqnarray}
b\kappa^2 \lambda^4= b \kappa_{n}^2(\lambda^4|_{\kappa_{n-1},b=0})
+\mathcal{O}(b^2)+\mathcal{O}[(\Delta\kappa)^4],
\end{eqnarray}
where we have assumed $\kappa_{-1}=0$,
$\lambda^2|_{\kappa_{-1}}=0$ and $\lambda^2|_{b=0}$ is the value
of $\lambda^2$ for $b=0$. Now we are going to compute the critical
temperature $T_{c}$. First of all, we start with the following
equation
\begin{eqnarray}\label{Tct}
T_{c}=\frac{f^{\prime} (r_{+c})}{4 \pi}.
\end{eqnarray}
From Eq. (\ref{f+}), $f^{\prime}(r_{+c})$ is expressed as
 \begin{eqnarray}
f^{\prime}(r_{+c})=3 r_{+c}-\kappa^2 r_{+c}
\left[\frac{{{\phi_{0}}^{\prime}}^{2}(r_{+c})}{2}+\frac{3}{2}b{{\phi_{0}}^{\prime}}^{4}(r_{+c}) \right].
\end{eqnarray}
Substituting Eq. (\ref{Phi}) in the above equation, and then
inserting the result back into Eq. (\ref{Tct}), we arrive at the
following expression for the critical temperature,
\begin{eqnarray}\label{Tc}
T_{c}=\frac{1}{4\pi} \sqrt{\frac{\rho}{\lambda}}\Bigg[ 3-
\frac{\kappa_{n} ^2 (\lambda^2|_{\kappa_{n-1}})}{2}+ \frac{1}{2} b
\kappa_{n}^2(\lambda^4|_{\kappa_{n-1},b=0}) \Bigg].
\end{eqnarray}
With these obtained computations out of the analytical approach at
hand, we are in a position to present the results of the critical
temperature $T_{c}$ for a $(2+1)$-dimensional holographic
superconductors in the presence of both EN electrodynamics as well
as backreaction. To do this, we assume the nonlinear parameter $b$
is small, by choosing it as $b = 0, 0.1, 0.2, 0.3$. We also get
the values $m^2=-2$, $\Delta_{i}=\Delta_{-}=1$ and
$\Delta\kappa=0.05$. As an example, we bring the details of our
calculations for the case of $n=4$ and summarize all results in
table $1$.

For $b=0$, we obtain $\lambda^2$ From Eq. (\ref{lambda2}) as
\begin{eqnarray}
\lambda^2=\frac{-8.279205 \alpha^2+4.924220
\alpha-4.957900}{-3.579048+0.878281 \alpha-0.153258 \alpha^2}.
\end{eqnarray}
From it we get the minimum eigenvalues of $\lambda^2$ and the
corresponding value of $\alpha$, as $\lambda^2_{\rm min}=1.2593$
at $\alpha=0.2361$. And thus the critical temperature is obtained
from Eq. (\ref{Tc}) as $T_{c}=0.2235 \sqrt{\rho}$, which is in
good agreement with the result of \cite{PJWC2012}.

For $b=0.1$, we find
\begin{eqnarray}
\lambda^2=\frac{-4.95286+4.91624\alpha-8.27411\alpha^2}{-3.035+0.673477\alpha-0.109325\alpha^2},
\end{eqnarray}
which has a minimum value $\lambda^2_{\rm min}=1.4757$ at
$\alpha=0.2417$, and we can get the critical temperature
$T_{c}=0.2147 \sqrt{\rho}$.

For $b=0.2$, we arrive at
\begin{eqnarray}
\lambda^2=\frac{-4.94387+4.90127\alpha-8.26415\alpha^2}
{-2.49088+0.468581\alpha-0.0652154\alpha^2},
\end{eqnarray}
whose minimum is $\lambda^2_{\rm min}=1.7811$ at $\alpha=0.24955$
and the critical temperature becomes $T_{c}=0.2046 \sqrt{\rho}$.

For $b=0.3$, we have
\begin{eqnarray}
\lambda^2=\frac{-4.93051+4.87819\alpha-8.24877\alpha^2}{-1.94661+0.263502\alpha-0.021046\alpha^2},
\end{eqnarray}
which attains its minimum $\lambda^2_{\rm min}=2.2451$ at
$\alpha=0.26133$ and the critical temperature reads $T_{c}=0.1927
\sqrt{\rho}$. We summarize our results for the critical
temperature in cases of different values of nonlinear and
backreaction parameters In table $1$. From this table, we see
that, for fixed value of the backreaction parameter, with the
nonlinear parameter $b$ getting stronger, the critical temperature
decreases. Similarly, for a fixed value of the nonlinear parameter
$b$, the critical temperature drops as the backreaction parameter
increases. Thus, we conclude that the critical temperature becomes
smaller and so, make the condensation harder when we increase the
values of both backreaction and nonlinear parameters. These
features were also observed in the study a $(2+1)$-dimensional
holographic superconductors with backreaction when the gauge field
is in the form of BI nonlinear electrodynamics \cite{BIBR}.
Comparing the results obtained here with those of \cite{BIBR}, we
observe that the effect of the EN corrections on the condensation
with respect to the BI nonlinear one is stronger when the
backreactions is taken into account in both cases. In other words,
the formation of scalar hair in the presence of EN electrodynamics
is harder compared to the case of BI nonlinear electrodynamics.
Obviously, our analytic results back up the findings in other
articles. In the case of $b=\kappa=0$, we observe that the
analytic results for the critical temperature are consistent with
both the analytical results of Ref. \cite{P.ZGJZ} as well as the
numerical result of Ref. \cite{Har2}. Also, we confirm the
numerical result found in Ref. \cite{ZPCJ.P.N.N} when the
backreaction parameter $\kappa$ is equal to zero. On the other
hand, for $b=0$ the data obtained for the critical temperature, is
analogous to those reported for the holographic superconductors
with backreaction in Maxwell theory \cite{PJWC2012}.
\begin{center}
\begin{tabular}{|c|cc|cc|cc|cc|}
\hline n & b=0 &  &b=0.1&  & b=0.2 &  & b=0.3 &  \\ \cline{2-9} &
BI & \multicolumn{1}{|c|}{EN} & BI & \multicolumn{1}{|c|}{EN} & BI
& \multicolumn{1}{|c|}{EN} & BI & \multicolumn{1}{|c|}{EN} \\
\hline \multicolumn{1}{|c|}{0} &0.2250  &
\multicolumn{1}{|c|}{0.2250} &0.2228  &
\multicolumn{1}{|c|}{0.2161} & 0.2206 &
\multicolumn{1}{|c|}{0.2060} & 0.2184 &
\multicolumn{1}{|c|}{0.1942 } \\ \hline 1&0.2249  &
\multicolumn{1}{|c|}{0.2249} &0.2227  &
\multicolumn{1}{|c|}{0.2160} & 0.2204 &
\multicolumn{1}{|c|}{0.2059} &0.2181  &
\multicolumn{1}{|c|}{0.1941} \\ \hline \multicolumn{1}{|c|}{2}
&0.2246  & \multicolumn{1}{|c|}{0.2246} & 0.2225 &
\multicolumn{1}{|c|}{0.2158} &0.2203  &
\multicolumn{1}{|c|}{0.2057} & 0.2180 &
\multicolumn{1}{|c|}{0.1938 } \\ \hline
 \multicolumn{1}{|c|}{3} &0.2241  & \multicolumn{1}{|c|}{0.2241} & 0.2220 &
\multicolumn{1}{|c|}{0.2153} & 0.2199 & \multicolumn{1}{|c|}{0.2050} & 0.2176 & \multicolumn{1}{|c|}{0.1934
} \\ \hline
 \multicolumn{1}{|c|}{4} & 0.2235 & \multicolumn{1}{|c|}{0.2235} & 0.2214 &
\multicolumn{1}{|c|}{0.2147} & 0.2192 & \multicolumn{1}{|c|}{0.2046} & 0.2170 & \multicolumn{1}{|c|}{0.1927
} \\ \hline
 \multicolumn{1}{|c|}{5} & 0.2226 & \multicolumn{1}{|c|}{0.2226} & 0.2208 &
\multicolumn{1}{|c|}{0.2141} & 0.2184 & \multicolumn{1}{|c|}{0.2038} & 0.2162 & \multicolumn{1}{|c|}{0.1919
} \\ \hline
 \multicolumn{1}{|c|}{6} & 0.2216 & \multicolumn{1}{|c|}{0.2216} & 0.2196 &
\multicolumn{1}{|c|}{0.2130} & 0.2174 & \multicolumn{1}{|c|}{0.2029} & 0.2152 & \multicolumn{1}{|c|}{0.1909
} \\ \hline
\end{tabular}
\\[0pt]
Table $1$: The critical temperature $T_{c}/\sqrt{\rho}$ for
holographic superconductors in the presence of BI and EN
electrodynamics. Here we have taken $\kappa_{n}=n \Delta \kappa$
where $\Delta \kappa=0.05$. The results for BI case are invoked
from Ref. \cite{BIBR}. \label{tab1}
\end{center}
%%%%%%%%%%%%%%%%%%%%%%%%%%%%%%%%%%%%%%%%%%%%%%%%%%%%%%%%%%%%%%%%%%%%%%%%%%
\section{CRITICAL EXPONENT AND THE CONDENSATION OF THE SCALAR OPERATOR}\label{Cri}
We use the Sturm-Liouville method to analytically examine the
scalar condensation and the order of the phase transition with
backreactions near the critical temperature. With the help of Eq.
(\ref{psi1F}), when $T$ is close to $T_{c}$, the equation of
motion (\ref{Phiz}) can be rewritten as
\begin{eqnarray}\label{phiCT}
\phi^{\prime\prime}\left(1+\frac{4bz^4}{r_{+}^2}\phi'^2\right) +
\frac{8b z^{3} }{r_{+} ^{2}} {\phi^{\prime}}^{3}= \frac{\langle
\mathcal{O} \rangle ^2}{r_{+} ^2}  \mathcal{B} (z) \phi (z),
\end{eqnarray}
\begin{eqnarray}
\mathcal{B} (z)= \frac{F^2 (z)}{1-z^3}
\left(1-\frac{2bz^4}{r_{+}^2} \phi'^2 (z) \right) \left[
1+\frac{\kappa^2 z^3}{1+z+z^2}
\left(\frac{\lambda^2}{2}-\frac{b\lambda^4}{10}
\xi(z)\right)\right],
\end{eqnarray}
where $\zeta(z)=1+z+z^2+z^3+z^4$. Since the parameter ${\langle
\mathcal{O} \rangle ^2}/{r_{+} ^2}$ is very small, we can expand
$\phi(z)$ as
\begin{eqnarray}\label{phibast}
\frac{\phi(z)}{r_{+}}=\lambda (1-z) \left(
1-\frac{2}{5}b\lambda^2\xi(z) \right)+\frac{\langle \mathcal{O}
\rangle ^2}{r_{+} ^2} \chi (z).
\end{eqnarray}
Substituting Eq. (\ref{phibast}) into Eq. (\ref{phiCT}), we can
obtain the equation of motion for $\chi(z)$ as
\begin{eqnarray}\label{dxhi}
\Bigg[K(z)\chi'(z)\Bigg]'&=& (1+4b\lambda^2
z^4)^{1/2}\frac{\lambda
F^2}{1+z+z^2} \nonumber \\
&& \times \Bigg\{1-\frac{2}{5}b\lambda^2(\xi(z)+5
z^4)+\frac{z^3}{1+z+z^2}\left(\frac{\kappa^2 \lambda^2}{2} -
\frac{b \kappa^2 \lambda^4}{10} (3\xi(z)+10z^4) \right) \Bigg\}.
\end{eqnarray}
with $\chi(1)=0=\chi'(1)$ and we have defined
\begin{eqnarray}
K(z)=\left(1+4b\lambda^2 z^4\right)^{3/2}.
\end{eqnarray}
Integrating both sides of Eq. (\ref{dxhi}) between $z=0$ to $z=1$,
we reach
\begin{eqnarray}\label{chi0}
\chi'(0)&=&-\lambda \int_{0}^{1} dz (1+4b\lambda^2 z^4)^{1/2}
\frac{F^2}{1+z+z^2} \nonumber \\
&& \times \Bigg\{1-\frac{2}{5}b\lambda^2(\xi(z)+5
z^4)+\frac{z^3}{1+z+z^2}\left(\frac{\kappa^2 \lambda^2}{2} -
\frac{b \kappa^2 \lambda^4}{10} (3\xi(z)+10z^4) \right) \Bigg\}.
\end{eqnarray}
Equating $\phi(z)$ from Eqs. (\ref{BC.Phi}) and (\ref{phibast}),
we arrive at
\begin{eqnarray} \label{murho}
\frac{\mu}{r_{+}}-\frac{\rho}{r_{+} ^2} z &=& \lambda (1-z)
\Bigg\{ 1-\frac{2}{5}b\lambda^2 \xi(z) \Bigg\}+\frac{\langle
\mathcal{O} \rangle ^2}{r_{+} ^2} \chi (z)\nonumber\\&=& \lambda
(1-z) \Bigg\{ 1-\frac{2}{5}b\lambda^2 \xi(z) \Bigg\}+\frac{\langle
\mathcal{O} \rangle ^2}{r_{+} ^2} \left(\chi(0)+z
\chi'(0)+...\right),
\end{eqnarray}
where in the last step we have expanded $\chi (z)$ around $z=0$.
Considering the coefficients of $z$ term in both sides of Eq.
(\ref{murho}), we find that
\begin{eqnarray}
\frac{\rho}{r_{+} ^2}=\lambda-\frac{\langle \mathcal{O} \rangle
^2}{r_{+} ^2}\chi'(0).
\end{eqnarray}
Substituting $\chi'(0)$  from Eq. (\ref{chi0}) in the above
relation, we get
\begin{eqnarray}\label{edit.CE}
\frac{\rho}{r_{+} ^2}=\lambda \Bigg\{ 1+\frac{\langle \mathcal{O}
\rangle ^2}{r_{+} ^2} \mathcal{A}\Bigg\},
\end{eqnarray}
where
\begin{eqnarray}
\mathcal{A}&=& \int_{0}^{1} dz (1+4b\lambda^2 z^4)^{1/2}
\frac{F^2}{1+z+z^2}\nonumber \\
&& \times \Bigg\{1-\frac{2}{5}b\lambda^2(\xi(z)+5
z^4)+\frac{z^3}{1+z+z^2}\left(\frac{\kappa^2 \lambda^2}{2} -
\frac{b \kappa^2 \lambda^4}{10} (3\xi(z)+10z^4) \right) \Bigg\}.
\end{eqnarray}
Using Eqs. (\ref{T}), (\ref{f+}) and (\ref{Phi}), and taking into
account the fact that $T$ is very close to $T_{c}$, we can deduce
\begin{eqnarray}\label{r+}
r_{+}=\frac{4\pi T}{\left[3-\frac{\kappa^2
\lambda^2}{2}+\frac{b}{2} \kappa^2 \lambda^4\right]}.
\end{eqnarray}
Eqs. (\ref{lambda}) and (\ref{r+}) show that Eq. (\ref{edit.CE}) can be rewritten as
\begin{eqnarray}
T_{c}^{2}-T^2= \langle \mathcal{O} \rangle ^2
\frac{\mathcal{A}}{(4\pi)^2} \left[3-\frac{\kappa^2
\lambda^2}{2}+\frac{b}{2} \kappa^2 \lambda^4 \right]^2.
\end{eqnarray}
Thus, we find the expectation value $\langle \mathcal{O} \rangle$
near the critical point as
\begin{eqnarray}\label{CE}
\langle \mathcal{O} \rangle=\gamma T_{c} \sqrt{1-\frac{T}{T_{c}}},
\end{eqnarray}
where $\gamma$ is the condensation parameter of the system which
is given by
\begin{eqnarray}
\gamma=\frac{4\pi\sqrt{2}}{\sqrt{\mathcal{A}}}\left[3-\frac{\kappa^2
\lambda^2}{2}+\frac{b}{2} \kappa^2 \lambda^4 \right]^{-1}.
\end{eqnarray}
The relation obtained in Eq. (\ref{CE}) is valid for small
nonlinear coupling and backreaction parameters and satisfies
$\langle \mathcal{O} \rangle \sim \sqrt{1-\frac{T}{T_{c}}}$.
Therefore, the analytical result supports that the phase
transition for the superconductor belongs to the second order and
the critical exponent of the system takes the mean-field value
$1/2$. This implies that considering nonlinear coupling and
backreaction parameters the value of the critical exponent will
not be altered. As we see in table $2$, condensation values
$\gamma$ increases with increasing the nonlinear parameter $b$ for
the fixed parameter $\kappa$. Also, we see the same behavior
between the condensation values $\gamma$ and the backreaction
parameter with a fixed value of the nonlinear parameter $b$. This
means that the condensation becomes harder to be formed by
considering both the nonlinear corrections to the gauge field and
taking the backreactions into account. It should be noted that, at
a temperature slightly below $T_{c}$ for the $(2+1)$-dimensional
holographic superconductors with backreaction, condensation values
for both BI and EN holographic superconductors have the same
behaviour, as we see in table $2$. Also, because of the larger
parameter $\gamma$, effect of the EN electrodynamics on the
condensation of the scalar operators is bigger than that of BI
case. This implies that the scalar hair is more difficult to be
developed in the holographic superconductors with EN
electrodynamics.
\begin{center}
\begin{tabular}{|c|cc|cc|cc|cc|}
\hline n & b=0 &  &b=0.1  &  & b=0.2 &  & b=0.3 &  \\ \cline{2-9}
& BI & \multicolumn{1}{|c|}{EN} & BI & \multicolumn{1}{|c|}{EN} &
BI & \multicolumn{1}{|c|}{EN} & BI & \multicolumn{1}{|c|}{EN} \\
\hline \multicolumn{1}{|c|}{0} &8.07  & \multicolumn{1}{|c|}{8.07}
&8.1801 & \multicolumn{1}{|c|}{8.5298} & 8.3094 &
\multicolumn{1}{|c|}{9.1579} & 8.4696 &
\multicolumn{1}{|c|}{10.0355 } \\ \hline 1&8.09  &
\multicolumn{1}{|c|}{8.09 } &8.1869  &
\multicolumn{1}{|c|}{8.5331} & 8.3212 &
\multicolumn{1}{|c|}{9.1616} &8.4890&
\multicolumn{1}{|c|}{10.0399} \\ \hline \multicolumn{1}{|c|}{2}
&8.11& \multicolumn{1}{|c|}{8.11} &8.1943&
\multicolumn{1}{|c|}{8.5443} &8.3237& \multicolumn{1}{|c|}{9.1742}
&8.4893& \multicolumn{1}{|c|}{10.0565 } \\ \hline
 \multicolumn{1}{|c|}{3} &8.115& \multicolumn{1}{|c|}{8.115} &8.2121&
\multicolumn{1}{|c|}{8.5630} &8.3417& \multicolumn{1}{|c|}{9.1951} &8.5023& \multicolumn{1}{|c|}{10.0818
} \\ \hline
 \multicolumn{1}{|c|}{4} & 8.13 & \multicolumn{1}{|c|}{8.13} & 8.2370 &
\multicolumn{1}{|c|}{8.5889} & 8.3669 &
\multicolumn{1}{|c|}{9.2241} & 8.5277 &
\multicolumn{1}{|c|}{10.1014 } \\ \hline
 \multicolumn{1}{|c|}{5} & 8.16 & \multicolumn{1}{|c|}{8.16} & 8.2909 &
\multicolumn{1}{|c|}{8.6425} & 8.3994 &
\multicolumn{1}{|c|}{9.2615} & 8.5606 &
\multicolumn{1}{|c|}{10.1639 } \\ \hline
 \multicolumn{1}{|c|}{6} & 8.20 & \multicolumn{1}{|c|}{8.20} & 8.3079 &
\multicolumn{1}{|c|}{8.6617} & 8.4391 &
\multicolumn{1}{|c|}{9.3070} & 8.6007 &
\multicolumn{1}{|c|}{10.2193 } \\ \hline
\end{tabular}
\\[0pt]
Table 2: The values of the condensation parameter $\gamma$ for
holographic superconductors in the presence of EN electrodynamics.
Here we have taken $\kappa_{n}=n \Delta \kappa$ where $\Delta
\kappa=0.05$. We have also provided the results for BI holographic
superconductor from Ref. \cite{BIBR}, for comparison. \label{tab2}
\end{center}
%%%%%%%%%%%%%%%%%%%%%%%%%%%%%%%%%%%%%%%%%%%%%%%%%%%%%%%%%%%%%%%%%%%
\section{Conclusions\label{Con}}
We have introduced a different type of gravity dual models, i.e.,
the charged AdS black holes in the context of Einstein-nonlinear
electrodynamics with a scalar field. We have assumed the EN
electrodynamics as the gauge field, and analytically investigated
the behavior of the $(2+1)$-dimensional holographic
superconductors. We have worked in a limit in which the scalar and
gauge fields backreact on the background metric. We have employed
the Sturm-Liouville analytic method to explore the problem. We
have found the influence of the nonlinear corrections to the gauge
filed as well as the backreaction effects on the critical
temperature and the process of the scalar field condensation. We
observed that the formation of the scalar hair condensation on the
boundary becomes harder in the presence of nonlinear
electrodynamics. This is mainly caused by the decreasing of the
critical temperature when the both nonlinear and backreaction
parameters become stronger. This phenomenon was also obtained in
the study of the effect of the BI and backreaction parameters in
the $(2+1)$-dimensional holographic superconductors \cite{BIBR}.
Comparing these different models show that for a specific $b$ the
critical temperature $T_{c}$ becomes larger for a holographic
superconductor with BI nonlinear electrodynamics comparing to the
case with EN electrodynamics. This implies that the scalar hair is
more difficult to develop in the latter case than the former one.
We have also given the critical exponent for the EN holographic
superconductor model with backreaction, which still takes the
mean-field value $1/2$. We found out that the condensation
parameter $\gamma$ in Eq. (\ref{CE}) increases with (i) increasing
the nonlinear parameter $b$ with a fixed backreaction parameter
$\kappa$, (ii) increasing the backreaction parameter with a fixed
value of the nonlinear parameter $b$. This implies that both the
nonlinear corrections to the gauge field as well as backreaction,
cause the formation of condensation harder.
%%%%%%%%%%%%%%%%%%%%%%%%%%%%%%%%%%%%%%%%%%%%%%%%%%%%%%%%%%%%%%%%%%%%%%%
\acknowledgments{We thank Shiraz University Research Council. This
work has been supported financially by Research Institute for
Astronomy and Astrophysics of Maragha (RIAAM), Iran.}
%%%%%%%%%%%%%%%%%%%%%%%%%%%%%%%%%%%%%%%%%%%%%%%%%%%%%%%%%%%%%%%%%%%%%%%

\end{document}